\documentclass[reprint,aps,prl,floatfix]{revtex4-2}
\usepackage{graphicx,amsmath,hyperref,pdfpages}
\makeatletter
\AtBeginDocument{\let\LS@rot\@undefined}
\makeatother

\begin{document}


\title{Many-body delocalisation as symmetry breaking}
\begin{abstract}
We present a framework in which the transition between a many-body localised (MBL) phase and an ergodic one is symmetry breaking. We consider random Floquet spin chains, expressing their averaged spectral form factor (SFF) as a function of time in terms of a transfer matrix that acts in the space direction. The SFF is determined by the leading eigenvalues of this transfer matrix. In the MBL phase the leading eigenvalue is unique, as in a symmetry-unbroken phase, while in the ergodic phase and at late times the leading eigenvalues are asymptotically degenerate, as in a system with degenerate symmetry-breaking phases. We identify the broken symmetry of the transfer matrix, introduce a local order parameter for the transition, and show that the associated correlation functions are long-ranged only in the ergodic phase.
\end{abstract}

\author{S. J. Garratt and J. T. Chalker}
\affiliation{Theoretical Physics, University of Oxford, Parks Road, Oxford OX1 3PU, United Kingdom}

\date{\today}

\maketitle

Broken symmetry has been a useful concept in the description of many phase transitions 
\cite{landau1980statistical,sachdev2011quantum,altland2010condensed}, extending even to the Anderson localisation of single-particle wavefunctions. In that context, it was a consideration of disorder-averaged Green's functions which showed how the corresponding delocalisation transition could be understood as symmetry breaking in a field theory \cite{wegner1979mobility}. There has since been substantial progress in the study of many-body localisation, the analogue of Anderson localisation in the presence of interactions \cite{fleishman1980interactions,gornyi2005interacting,basko2006metal,oganesyan2007localization,pal2010many,nandkishore2015manybody,abanin2019colloquium}. The many-body localisation transition, however, has not so far been described in the language of symmetry breaking, nor has a local order parameter been identified.

The many-body localised (MBL) phase is characterised by its failure to equilibrate, and so falls outside the regimes described by statistical mechanics \cite{nandkishore2015manybody,abanin2019colloquium}. It is to be contrasted with the ergodic phase, where in the thermodynamic limit local observables approach their equilibrium values under unitary dynamics \cite{dallesio2016from,deutsch2018eigenstate,deutsch1991quantum,srednicki1994chaos,rigol2008thermalization}. 
This fundamental difference in dynamics is reflected in the spectral properties; indeed the many-body localisation transition, which separates the MBL and ergodic phases, can be characterised as an eigenstate phase transition~\cite{nandkishore2015manybody}.

Because the dynamics is in question it is natural to adopt a space-time description. Such an approach has been central to recent advances in our understanding of entanglement growth \cite{nahum2017quantum,keyserlingk2018operator,nahum2018operator} and the measurement transition \cite{li2018quantum,skinner2019measurement,chan2019unitary,li2019measurement} in many-body systems. In these investigations random quantum circuits, minimal models for discrete time evolution with local interactions, have proved exceedingly useful. In addition, their time-periodic counterparts, random Floquet circuits, have cast light on the spectral properties of local evolution operators \cite{kos2018analytic,chan2018solution,chan2018spectral,bertini2018exact,garratt2020manybody}, including examples which are believed to be representative of Floquet systems more generally.
 
\begin{figure}
\includegraphics[width=0.4\textwidth]{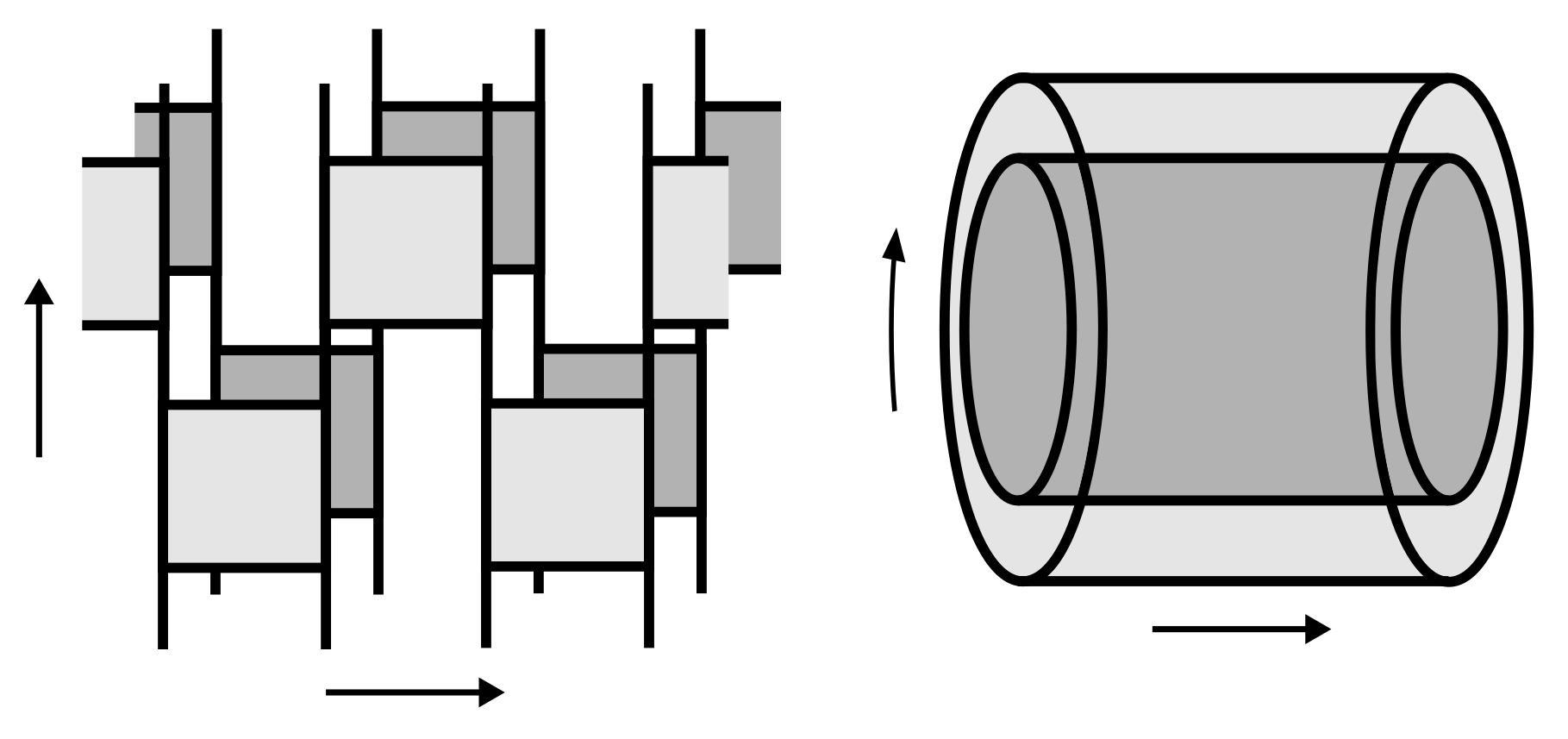}
\put(-208,80){(a)}
\put(-97,80){(b)}
\put(-154,-2){$x$}
\put(-208,46){$t$}
\put(-46,6){$x$}
\put(-97,53){$t$}
\put(-52,51){$K(t)$}
\put(-180,30){$U_{0,1}$}
\put(-138,30){$U_{2,3}$}
\put(-159,62){$U_{1,2}$}
\vspace{0.1in}

\includegraphics[width=0.4\textwidth]{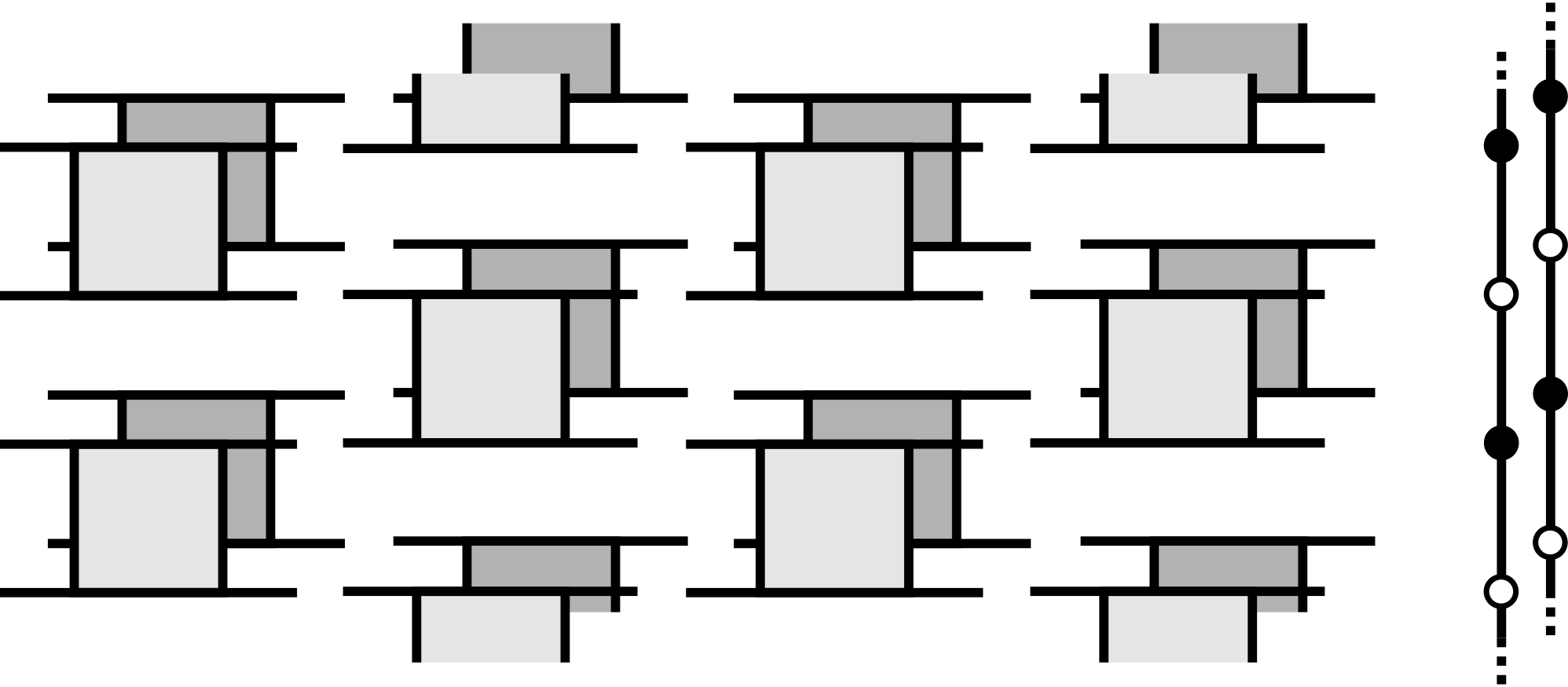}
\put(-208,90){(c)}
\put(-191,87){$\mathcal{R}_{0,1}$}
\put(-150,93){$\mathcal{S}\mathcal{R}_{1,2}\mathcal{S}^{-1}$}
\put(-100,87){$\mathcal{R}_{2,3}$}
\put(-61,93){$\mathcal{S}\mathcal{R}_{3,4}\mathcal{S}^{-1}$}
\put(-20,9){$a_0$}
\put(3,17){$a^*_0$}
\put(-20,29){$b_0$}
\put(3,37){$b^*_0$}
\put(-20,49){$a_1$}
\put(3,57){$a^*_1$}
\put(-20,67){$b_1$}
\put(3,75){$b^*_1$}
\caption{(a) Floquet operator $W$ (light) and its complex conjugate $W^*$ (dark) with space $x$ running horizontally and time $t$ running vertically. Vertical black lines represent the positions of spins, and squares the $4 \times 4$ unitary matrices $U$ which act on nearest neighbours. (b) Illustration of the spectral form factor $K(t)$. The light outer cylinder represents $\text{Tr}W(t)$ and the dark inner cylinder $\text{Tr}W^*(t)$. (c) Matrices acting on the space of single-site orbit pairs, shown as the $2 \times (2t)$-site spin ladder on the right, to generate $K(t)$.}
\label{fig:circuit}
\end{figure}
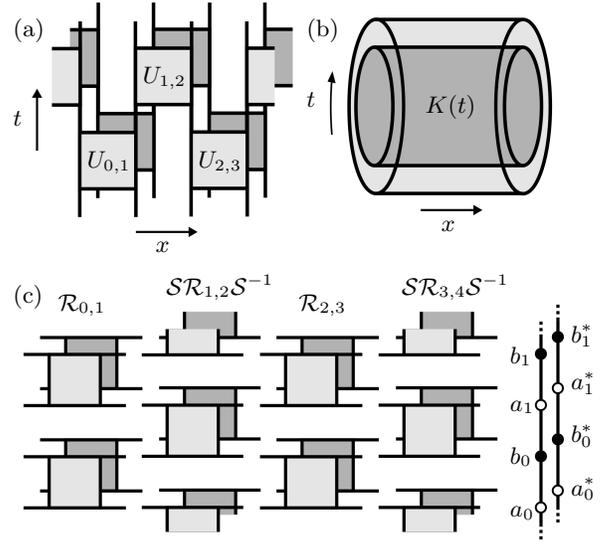 
 
In this Letter we study the spectral form factors (SFFs) of random Floquet circuits across a many-body localisation transition \cite{ponte2015manybody,lazarides2015fate,zhang2016floquet,sunderhauf2018localization,mace2019quantum}. The SFF $K(t)$ at time $t$ probes spectral statistics on quasienergy scales $2\pi/t$, and its disorder average ${\langle K(t) \rangle}$ is very different in the two phases. For a spin-1/2 chain with $L$ sites, and without time-reversal symmetry, ${\langle K(t)\rangle} \simeq t$ in the ergodic phase for sufficiently large $t < 2^L \equiv t_{\text{H}}$, the Heisenberg time. On the other hand, in the MBL phase ${\langle K(t) \rangle} \simeq 2^L$ for $t \gtrsim 2$. Our approach, as in Ref.~\cite{garratt2020manybody}, is to express the disorder-averaged SFF in terms of an averaged transfer matrix which acts in the space direction.

Related transfer matrices have been used to study kicked Ising models \cite{akila2016particle,akila2017semiclassicalprl,akila2018semiclassicalannals}, while those generating the SFF have been used to study the dual-unitary point \cite{bertini2018exact,flack2020statistics} and signatures of localisation away from it \cite{braun2020transition}. Analogous objects have been introduced to study entanglement growth and dynamical correlations in dual-unitary models \cite{bertini2019entanglement,bertini2020exact}, as well as thermalisation via an influence matrix in more general settings \cite{lerose2020influence,sonner2020characterizing}. Using the transfer matrix representation of the average SFF we will show that the MBL-ergodic transition is symmetry breaking.

Our models are Floquet spin-1/2 Heisenberg chains with random local fields. The evolution operator for integer time $t$ is $W(t) \equiv W^t$, where the Floquet operator $W$ is a $2^L \times 2^L$ unitary matrix with the brickwork structure shown in Fig.~\ref{fig:circuit}(a). During a single time step each site $x=0 \ldots (L-1)$ interacts first with one neighbour and then with the other. These interactions are described by $4 \times 4$ unitary matrices $U_{x-1,x}$ and $U_{x,x+1}$ with the parametrisation
\begin{align}
	U = &[B \otimes B'] [\cos(\pi J) + i \sin (\pi J) \Sigma][A \otimes A'].
\label{eq:U}
\end{align}
Here $A,A',B$ and $B'$ are independent Haar-random $\text{U}(2)$ matrices representing the local fields, $\Sigma$ is the two-site swap operator, and $J$ is the coupling strength. Our model has no conserved densities, and does not have time-reversal symmetry. Furthermore, all points in the spectrum are statistically equivalent, precluding a mobility edge and removing the need for any unfolding procedure in the analysis of the spectrum. For $J=1/2$ our model is dual unitary, falling into the non-interacting class of Refs.~\cite{claeys2020ergodic,bertini2019exact}. Here, however, we are concerned only with ${0 \leq J \leq 0.3}$. Our model is MBL for $J < J_c$ and ergodic for $J_c < J < 1/2$, with the critical point $J_c \approx 0.07$ \cite{supplementary}. We focus on behaviour within each of the two phases and defer consideration of critical properties for future work.

The SFF is defined by $K(t) = |\text{Tr}W(t)|^2$ [see Fig.~\ref{fig:circuit}(b)], and using the spectral decomposition $W=\sum_n e^{i \theta_n}{|n\rangle \langle n|}$ with $n=1 \ldots 2^L$ we see that $K(t) = \sum_{nm}e^{i (\theta_n -\theta_m)t}$. This is the Fourier transform of the two-point correlator of the level density. Moreover, $\text{Tr}W(t)$ can be expressed as a sum over closed paths, or many-body orbits, in the space of spin configurations. $K(t)$ is then a sum over pairs of such many-body orbits; we will refer to those coming from $\text{Tr}W(t)$ as forward orbits, and those coming from the conjugate as backward orbits. For Floquet systems with local interactions this sum over pairs of forward and backward orbits can be generated by transfer matrices acting in the space direction, as we now discuss.

The orbits of individual spins can be represented as the states of a spin chain, and orbit pairs as the states of a spin ladder. To see this note that these orbits are sequences of $2t$ states, and we can write these as vectors $|a_0 b_0 \ldots a_{t-1} b_{t-1}\rangle$, where $a_r,b_r=0,1$ represent the state at times $r$ and $(r+1/2)$ respectively, for $r$ integer. Orbit pairs are then naturally represented as tensor products $|a_0 b_0 \ldots\rangle \otimes |a^*_0 b^*_0 \ldots\rangle$. Here unstarred labels correspond to the forward orbit, and starred to the backward. These are the states of a $2 \times (2t)$-site spin ladder, shown on the right in Fig.~\ref{fig:circuit}(c).

The transfer matrices generating $K(t)$ act on states of this spin ladder. In a given disorder realisation these are tensor products of matrices which act on the two spin chains, generating $\text{Tr}W(t)$ and its conjugate, respectively [see Fig.~\ref{fig:circuit}(c)]. Details of this construction are given in Refs.~\cite{garratt2020manybody,supplementary}. We denote matrices generating $K(t)$ in this way by $\mathcal{R}_{x,x+1}$ for bonds $(x,x+1)$ with $x$ even, and it is clear from Fig.~\ref{fig:circuit}(c) that for $x$ odd these matrices are shifted by half of a time step with respect to those with $x$ even. On single-site orbits this shift is described by the operator $S$ defined by $S|a_0 b_0 \ldots a_{t-1} b_{t-1}\rangle = |b_0 a_1 \ldots b_{t-1} a_0\rangle$. On orbit pairs, it is described by the operator $\mathcal{S}=S \otimes S$. 

With periodic boundary conditions, which necessitates $L$ even, we then have
\begin{align}
	K(t) = \text{tr}[ \mathcal{R}_{0,1} \mathcal{S} \mathcal{R}_{1,2} \mathcal{S}^{-1} \ldots  \mathcal{S} \mathcal{R}_{L-1,0} \mathcal{S}^{-1}],
\end{align}
where the trace $\text{tr}$ is over single-site orbit pairs. Because the matrices $U_{x,x+1}$ are independently and identically distributed, $\mathcal{R}_{x,x+1}$ can be averaged independently for each $x$ and we write ${\langle\mathcal{R}_{x,x+1}\rangle} = {\langle\mathcal{R}\rangle}$. The average SFF is then
\begin{align}
	{\langle K(t)\rangle} = \text{tr} [\mathcal{S}^{-L}[\mathcal{S}{\langle\mathcal{R}\rangle}]^L ] \label{eq:SFFpbc},
\end{align}
where we have used time periodicity to write $\langle K(t) \rangle$ in terms of a single kind of averaged transfer matrix, $\mathcal{S}\langle \mathcal{R} \rangle$. Open boundary conditions are instead encoded in vectors $\langle \mathcal{B}_L|$ and $|\mathcal{B}_R\rangle$ so that
\begin{align}
	{\langle K(t)\rangle} = \langle \mathcal{B}_L|[\mathcal{S}{\langle\mathcal{R}\rangle}]^{L-1}|\mathcal{B}_R \rangle.
\label{eq:SFFobc}
\end{align}

We now elaborate on how the MBL-ergodic transition is symmetry breaking. The transfer matrices, which act on pairs of single-site orbits, commute with the time translation operations $S^2 \otimes 1$ and $1 \otimes S^2$ acting on the respective forward and backward orbits. These symmetries imply that $\mathcal{S}{\langle\mathcal{R}\rangle}$ can be block-diagonalised into $t^2$ sectors labelled by the $t$ different eigenvalues of each of $S^2 \otimes 1$ and $1 \otimes S^2$. In the regimes we consider no more than $t$ eigenvalues control ${\langle K(t)\rangle}$, and one of these resides in each sector with eigenvalue $e^{2 \pi i \nu /t}$ under $S^2 \otimes 1$ and $e^{-2 \pi i \nu/t}$ under $1 \otimes S^2$, for integer $\nu=0 \ldots (t-1)$. We denote these eigenvalues of $\mathcal{S}{\langle \mathcal{R}\rangle}$ by $\lambda(\nu,t)$, and the corresponding right and left eigenvectors by $|\nu,t;R \rangle$ and $\langle \nu,t;L|$, respectively. In the MBL phase $\lambda(0,t)$ is the unique leading eigenvalue, and the corresponding eigenvector is invariant under relative translation of forward and backward orbits. In the ergodic phase all $\lambda(\nu,t)$ are asymptotically degenerate at large $t$. The corresponding eigenvectors break the symmetry of the transfer matrix under relative time-translation of orbits within a pair.

This behaviour of the eigenvalues can be related to the dependence of $\langle K(t) \rangle$ on $t$ and $L$ in the two phases as follows. First note that in the limit of decoupled sites $(J=0)$, $\langle K(t) \rangle = \langle k(t) \rangle^L$, where $\langle k(t) \rangle$ is the average SFF for $2 \times 2$ Haar-random unitary matrices. $\langle k(1)\rangle=1$ and $\langle k(t \geq 2) \rangle =2$ so $\langle K(t)\rangle$ saturates at $2^L$ for $t \geq 2$. In this case there is one nonzero eigenvalue $\lambda(0,t)=\langle k(t)\rangle$. For small $J<J_c$ the other eigenvalues are nonzero, but $\lambda(0,t)$ remains dominant. By contrast in the ergodic phase ${\langle K(t)\rangle} \simeq t$ for ${t_{\text{Th}}<t<t_{\text{H}}}$, where $t_{\text{Th}}$ (a function of $L$) is the Thouless time. This arises from having $t$ eigenvalues $\lambda(\nu,t) \simeq 1$ for all $\nu$ \cite{garratt2020manybody}.

\begin{figure}
\includegraphics[width=0.47\textwidth]{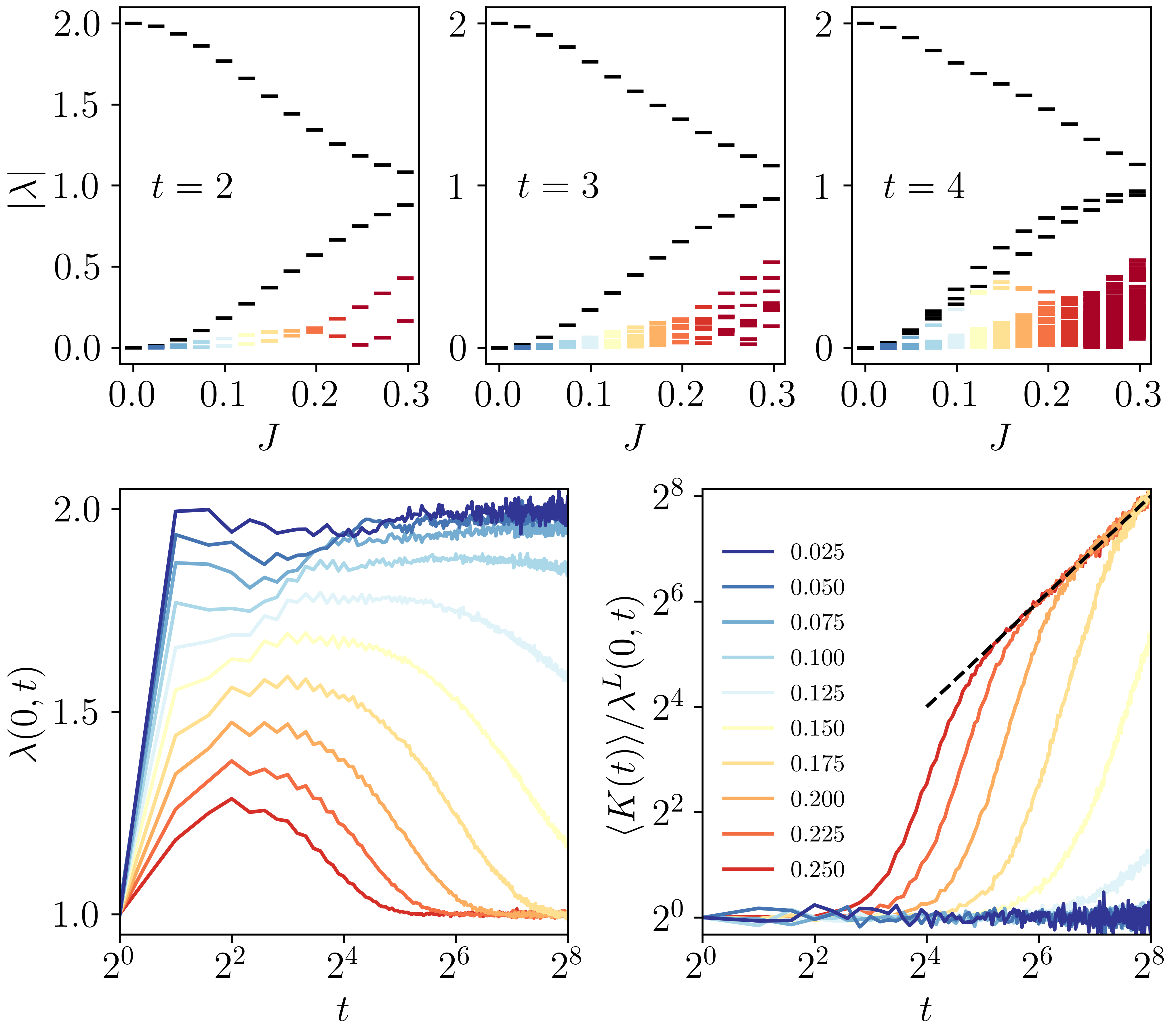}
\put(-167,202){(a)}
\put(-93,202){(b)}
\put(-17,202){(c)}
\put(-214,104){(d)}
\put(-95,104){(e)}
\caption{(a)-(c) Absolute values of all nonzero eigenvalues $\lambda$ of the transfer matrix $\mathcal{S}{\langle \mathcal{R}\rangle}$ at times (a) $t = 2$ (b) $t=3$ and (c) $t=4$, with $J=0.025m$ and $m=0,1 \ldots 12$. The $t$ leading eigenvalues at each $J$, some of which are degenerate, are shown black. For these times the largest is unique and in the $\nu=0$ sector. (d) $\lambda(0,t)$ calculated from the average SFF with open boundary conditions using Eq.~\eqref{eq:SFFobc_expansion}. (e) Ratio of ${\langle K(t) \rangle}$ with periodic boundary conditions, and with $L=12$, to the contribution of only the $\nu=0$ sector. The black dashed line shows $t$.}
\label{fig:lambda}
\end{figure}
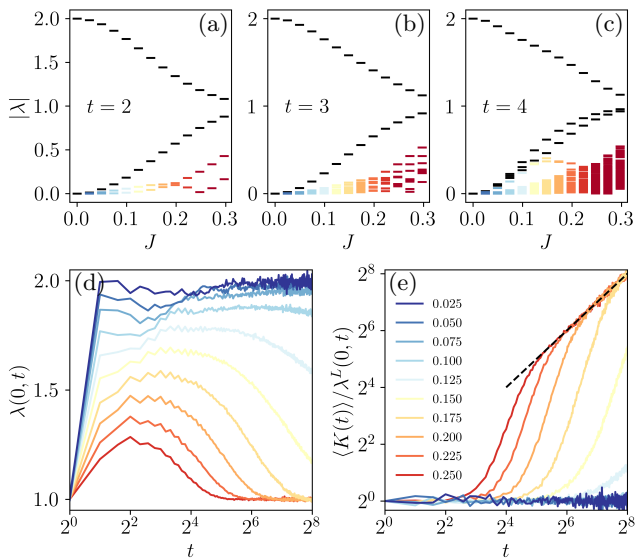

To support these claims and investigate behaviour at general $J$ we use a variety of numerical approaches. At very short times we can diagonalise $\mathcal{S} {\langle\mathcal{R}\rangle}$ exactly and in Figs.~\ref{fig:lambda}(a)-(c) we show the magnitudes of all nonzero eigenvalues for $0 \leq J \leq 0.3$ and $t=2,3,4$. Even for these small values of $t$ we see that increasing $J$ causes the separation between the $\lambda(\nu,t)$ to decrease, and a gap between them and all other eigenvalues to appear. For these times $\lambda(0,t)>1$, whereas $\lambda(\nu \neq 0,t)<1$, with $\lambda(0,t) \gg \lambda(\nu \neq 0, t)$ at small $J$.

To elucidate the behaviour of $\mathcal{S}{\langle\mathcal{R}\rangle}$ in each phase it is necessary to go to much larger $t$, where exact diagonalisation is computationally too demanding. Instead we compare the average SFF with open and periodic boundary conditions. Because $\langle \mathcal{B}_L|$ and $|\mathcal{B}_R\rangle$ are invariant under translation by integer time steps, the SFF with open boundary conditions has contributions from only the $\nu=0$ sector, while with periodic boundary conditions all sectors contribute. 

From Eq.~\eqref{eq:SFFobc} and the spectral decomposition of $\mathcal{S}{\langle\mathcal{R}\rangle}$ we have with open boundary conditions
\begin{align}
	{\langle K(t)\rangle} = \lambda^{L-1}(0,t)\langle \mathcal{B}_L|0,t;R \rangle \langle 0,t;L|\mathcal{B}_R \rangle + \ldots,
\label{eq:SFFobc_expansion}
\end{align}
where the ellipses represent the contributions from subleading eigenvalues. Note that here the result ${\langle K(t)\rangle} \simeq t$ at late times in the ergodic phase arises from $\lambda(0,t) \simeq 1$ and $\langle \mathcal{B}_L|0,t;R \rangle \langle 0,t;L|\mathcal{B}_R \rangle \simeq t$ \cite{garratt2020manybody}. From the scaling of ${\langle K(t)\rangle}$ with $L$ at fixed $t$ we extract $\lambda(0,t)$, and the results are shown in Fig.~\ref{fig:lambda}(d) for various $J$ and for $t \leq 2^8$. At late times we see for small $J$ that $\lambda(0,t) \simeq 2$, while for large $J$, $\lambda(0,t) \simeq 1$.

With periodic boundary conditions
\begin{align}
	{\langle K(t)\rangle} = \sum_{\nu=0}^{t-1} \lambda^L(\nu,t) + \ldots,
\label{eq:SFFpbc_expansion}
\end{align}
where again the ellipses represent contributions from subleading eigenvalues. For a transfer matrix $\mathcal{S}{\langle\mathcal{R}\rangle}$ with $t$ degenerate leading eigenvalues $\lambda(\nu,t)=1$, and all others negligible, we expect ${\langle K(t) \rangle}/\lambda^L(0,t)=t$. If only one eigenvalue dominates, and it is in the symmetric $\nu=0$ sector, ${\langle K(t) \rangle}/\lambda^L(0,t) = 1$. Precisely this behaviour, in the ergodic and MBL phases, respectively, is shown in Fig.~\ref{fig:lambda}(e).

To probe symmetry breaking more directly we introduce a local order parameter. The relevant symmetry, time-translation of backward orbits with respect to forward orbits, is at time $t$ the symmetry of a $t$-state clock model. We can therefore represent the order parameter by a complex number. We define a diagonal operator $\mathcal{C}$ acting on the space of single-site orbit pairs
\begin{align}
	\mathcal{C}&|a_0 b_0 \ldots\rangle \otimes |a_0^* b_0^* \ldots\rangle \\ &= \frac{2}{t}\sum_{p,r=0}^{t-1} \delta_{a_p a_r^*} e^{2\pi i(p-r)/t} |a_0 b_0 \ldots\rangle \otimes |a_0^* b_0^* \ldots\rangle. \notag
\end{align}
The diagonal matrix elements of $\mathcal{C}$ in this basis are the values of the order parameter for each orbit pair.

We want to study correlations of $\mathcal{C}$ in the averaged sum over orbit pairs that generates the SFF. The one-point function $\mathcal{O}_1(x,t)$ associated with our order parameter is obtained by inserting $\mathcal{C}$ into the sum at the site $x$, averaging, and dividing the result by ${\langle K(t) \rangle}$. For example, with periodic boundary conditions,
\begin{align}
	\mathcal{O}_1(x,t) = \frac{\text{tr}[\mathcal{S}^{-L}\mathcal{C}[\mathcal{S}\langle \mathcal{R}\rangle]^L]}{\text{tr}[\mathcal{S}^{-L}[\mathcal{S}\langle \mathcal{R}\rangle]^L]}.
\label{eq:onepointfunction}
\end{align}
Since our model has no symmetry-breaking fields or boundary conditions $\mathcal{O}_1(x,t)$ is identically zero. In Eq.~\eqref{eq:onepointfunction} this is due to a selection rule on matrix elements of $\mathcal{C}$ which arises from $(S^{2} \otimes 1) \mathcal{C} (S^{-2} \otimes 1) = e^{2\pi i/t}\mathcal{C}$. The two-point function $\mathcal{O}_2(x,y,t)$ is found by inserting the operators $\mathcal{C}^*$ and $\mathcal{C}$ at the sites $x$ and $y$, respectively. We expect $\mathcal{O}_2(x,y,t)$ to show long-range correlations in a symmetry-broken phase.

Using the spectral decomposition of the transfer matrix, and the selection rule indicated above, the two-point function in the thermodynamic limit $L\to\infty$ can be written
\begin{align}
	\mathcal{O}_2(x,y,t) &= [\lambda(1,t)/\lambda(0,t)]^{|x-y|}F(t) + \ldots,
\label{eq:twopointfunction}
\end{align}
where the form of the (real) amplitude $F(t)$ depends on the parity of $|x-y|$. For example, with $|x-y|$ even, $F(t)=\langle 0,t;L|\mathcal{C}^*|1,t;R\rangle\langle 1,t;L|\mathcal{C}|0,t;R\rangle$ \cite{supplementary}. The ellipses denote the contributions from subleading eigenvalues. From Eq.~\eqref{eq:twopointfunction} we find the correlation length $\xi(t)$ with
\begin{align}
	1/\xi(t)= \ln|\lambda(0,t)/\lambda(1,t)|.
\label{eq:correlationlength}
\end{align}
Hence $\xi(t)$ is small if the transfer matrix  has a single dominant eigenvalue $\lambda(0,t)$, as in the MBL phase, and is divergent if $\lambda(0,t)$ and $\lambda(1,t)$ are degenerate at large $t$, as in the ergodic phase.

Alternatively, using the spectral decomposition of the Floquet operator, the two-point function can be expressed as \cite{supplementary}
\begin{align}
	\mathcal{O}_2(x,y,t)= \big\langle \big|\sum_{mn} G_{mn}(t) \langle n|Z(x)|m \rangle \\ \times \langle m|Z(y)|n \rangle\big|^2 \big\rangle /{\langle K(t)\rangle} \notag
\end{align}
where $Z(x)$ is the Pauli matrix acting at site $x$, and $G_{mn}(t) = e^{i\theta_n t} \sum_{r=0}^{t-1} e^{i(2\pi/t + \theta_m - \theta_n)r}$ selects for quasienergy separations $(\theta_n-\theta_m) \sim 2\pi/t$. From this perspective, in the MBL phase we expect $\mathcal{O}_2(x,y,t)$ to be small for large $|x-y|$, because in that case there are few pairs of eigenstates $n$ and $m$ for which $\langle n|Z(x)|m \rangle$ and $\langle m|Z(y)|n \rangle$ are both large. Conversely, modelling these matrix elements using the eigenstate thermalisation hypothesis \cite{dallesio2016from,deutsch2018eigenstate,deutsch1991quantum,srednicki1994chaos,rigol2008thermalization}, expected to be applicable in the ergodic phase for $t \gg t_{\text{Th}}$, yields $\mathcal{O}_2(x,y,t)=1$ for $x \neq y$ and $t \ll t_{\text{H}}$ \cite{supplementary}.

\begin{figure}
\includegraphics[width=0.47\textwidth]{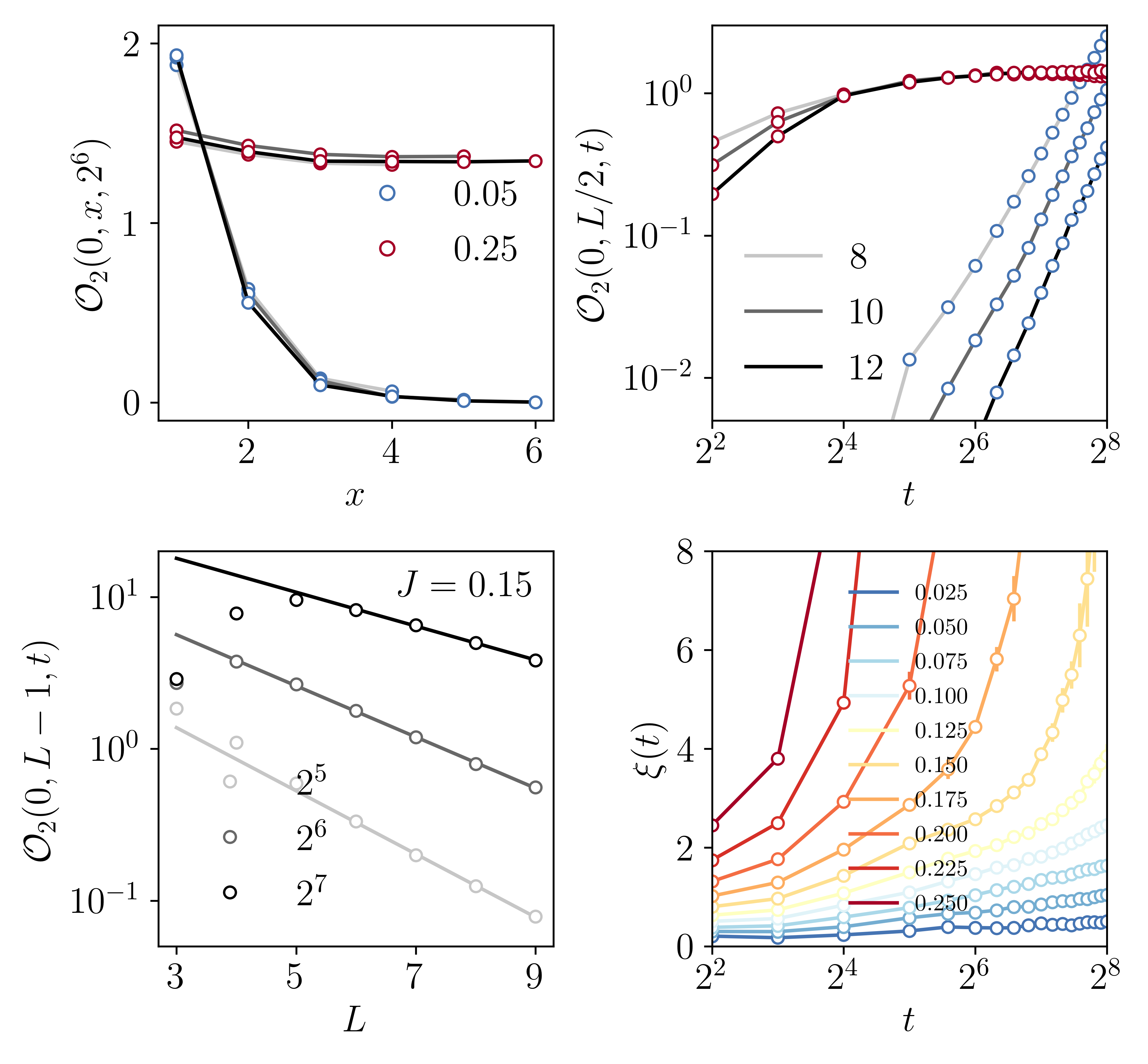}
\put(-206,209){(a)}
\put(-90,209){(b)}
\put(-206,99){(c)}
\put(-90,99){(d)}
\caption{Two-point function. With periodic boundary conditions, (a) $\mathcal{O}_2(0,x,t=2^6)$ versus $x$ and (b) $\mathcal{O}_2(0,x=L/2,t)$ versus $t$ in the MBL ($J=0.05$) and ergodic ($J=0.25$) phases. $L=8,10,12$ as indicated. (c) $L$-scaling of $\mathcal{O}_2(0,L-1,t)$ with open boundary conditions, with $t$ on the legend. Lines are fits to $L \geq 6$. (d) Correlation length $\xi(t)$ for various $J$, extracted from $\mathcal{O}_2(0,L-1,t)$ as in (c).}
\label{fig:pairing}
\end{figure}
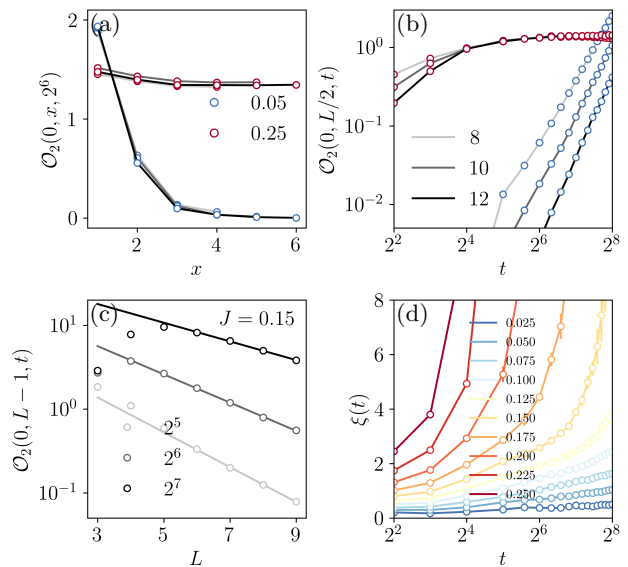

In Fig.~\ref{fig:pairing} we test these suggestions for the behaviour of $\mathcal{O}_2(x,y,t)$ against numerics. From Fig.~\ref{fig:pairing}(a) we see a rapid decay with $|x-y|$ in the MBL phase {($J=0.05$)}, but find $\mathcal{O}_2(x,y,t)$ approximately independent of $|x-y|$ in the ergodic phase {($J=0.25$)}. Aspects of this behaviour depend on $t$, as we examine in Fig.~\ref{fig:pairing}(b). In the ergodic phase, variation is weak provided $t$ is sufficiently large. By contrast, in the MBL phase the amplitude of the two-point function increases with $t$ as a power law, and $\xi(t)$ increases slowly but remains small over the accessible range of $t$. We note that at low frequencies both power-law growth of matrix elements of local operators, and logarithmic growth in an associated lengthscale, are known features of the MBL phase \cite{serbyn2016spectral,gopalakrishnan2015low,serbyn2017thouless,crowley2021constructive,garratt2021local}. To investigate in detail the dependence of $\xi(t)$ on $t$, we extract it from the scaling of $\mathcal{O}_2(x,y,t)$ with separation \cite{supplementary}, as illustrated in Fig.~\ref{fig:pairing}(c). Results for a range of $t$ and $J$ are shown in Fig.~\ref{fig:pairing}(d). These show a very rapid divergence of $\xi(t)$ with $t$ in the ergodic phase, and a much slower increase in the MBL phase.

The fact that $\xi(t)$ increases with $t$ for \emph{all} values of $J$ raises the question of whether it is possible to make a sharp distinction between the two phases from the behaviour of $\mathcal{O}_2(x,y,t)$. This can be answered in the affirmative using the links between (i) $\xi(t)$ and $\lambda(\nu,t)$ [Eq.~\eqref{eq:correlationlength}], and (ii) $\lambda(\nu,t)$ and $\langle K(t) \rangle$ [Eq.~\eqref{eq:SFFpbc_expansion}], together with the known behaviour of $\langle K(t) \rangle$ [Fig.~\ref{fig:lambda}]. These imply in the ergodic phase that $\xi(t) \gg L$ for $t_{\rm Th} \ll t < t_{\text{H}}$, since all $\lambda(\nu,t)$ are quasi-degenerate, and in the MBL phase that $\xi(t) \ll L$ for $t<t_{\text{H}}$, since $\lambda^L(0,t) \gg \lambda^L(\nu \neq 0,t)$. Hence, for large $L$ and $t$, $\mathcal{O}_2(x,y,t)$ reveals long-range order in the ergodic phase and disorder in the MBL phase. A prescription that ensures the two phases are distinguished is to take the limits $L,t\to \infty$ with $t=t_{\rm H}f \equiv 2^Lf$ for fixed $0<f<1$.

In summary we have shown how, in a spatially extended many-body Floquet system with local interactions, the transition from an MBL phase to an ergodic one can be viewed as symmetry breaking. To do this we have related the disorder-averaged SFF to the spectrum of a transfer matrix acting in the space direction. In the ergodic phase this transfer matrix has $t$ leading eigenvalues which are asymptotically degenerate at large $t$, and these are associated with symmetry-breaking eigenvectors. The MBL phase on the other hand is characterised by a transfer matrix with one dominant eigenvalue, and the corresponding eigenvector breaks no symmetries. We have defined a local order parameter for the transition, and have shown that the behaviour of the SFF in the ergodic phase is associated with long-range correlations of this order parameter.

This perspective on the many-body localisation transition, and the set of tools we have developed, open new opportunities and raise a number of questions. While our focus here has been on behaviour deep within each phase, one can ask, for example, how $\xi(t)$ behaves in the vicinity of the critical point. Further questions concern the roles played by time-reversal symmetry, and by charge conservation. Finally, it would of course be interesting to adapt our ideas to a Hamiltonian system. This will require technical developments paralleling work described for Floquet systems in Ref.~\cite{garratt2020manybody}; a possible route might build on recent work described in Ref.~\cite{winer2020hydrodynamic}. Note that the broken symmetry in the Hamiltonian setting would be a continuous one.

In contrast with work emphasising rare-region effects at the transition \cite{deroeck2017stability,agarwal2017rare}, here we have focused only on an averaged quantity, the SFF. This nevertheless captures the distinction between the MBL and ergodic phases. A natural next step is to consider higher moments of the SFF, whereas an alternative is to study products of unaveraged transfer matrices, and the associated Lyapunov exponents \cite{chan2020spectral}.

We are grateful to S. Roy for collaboration on related work, and to S. Parameswaran and A. Nahum for very useful discussions. This work was supported in part by EPSRC Grants EP/N01930X/1 and EP/S020527/1.

\bibliographystyle{apsrev4-2}
\bibliography{mblpairingrefs_arxiv2}

\clearpage


\section*{Supplementary material}

# Supplemental material for "Many-Body Delocalisation as Symmetry Breaking"

S. J. Garratt and J. T. Chalker

*Theoretical Physics, University of Oxford, Parks Road, Oxford OX1 3PU, United Kingdom*
(Dated: October 20, 2021)

In "Many-Body Delocalisation as Symmetry Breaking" we have studied the spectral form factors (SFFs) of a class of random Floquet spin chains across a many-body localisation transition. Before outlining the content of this note, we briefly give additional details on numerical calculations appearing in that Letter. Figs. 2(d) and (e) are based on length-scaling of the average SFF; SFFs were calculated via exact diagonalisation (ED) of Floquet operators, and averaged over $10^6$ disorder realisations for system sizes $3 \leq L \leq 8$, $2 \times 10^5$ for $L = 9, 10$, and $2 \times 10^4$ for $L = 11, 12$. In Fig. 3 we have calculated a certain averaged two-point function, also via ED of Floquet operators. There for $L \leq 9$ we average over $10^5$ disorder realisations, and for $L \geq 10$ over $10^4$.

Here we discuss the following technical aspects of the Letter

S1 Standard spectral and entanglement probes of the MBL transition in our model.

S2 Construction of the transfer matrix generating the SFF.

S3 Correlations of the clock-model order parameter.

S4 Response of the SFF to symmetry-breaking fields.

## S1. SPECTRAL AND ENTANGLEMENT PROBES OF THE TRANSITION

Here we characterise the MBL transition in our model. First, as in the main text, we consider the many-body spectrum. A standard probe is the $r$-statistic [S1], defined as follows. Writing $\Delta\theta_n = |\theta_{n+1} - \theta_n|$ with $\theta_n$ the ordered quasienergies of the Floquet operator, $r \equiv \min(\Delta\theta_n, \Delta\theta_{n+1})/\max(\Delta\theta_n, \Delta\theta_{n+1})$. In the ergodic phase we expect $r \simeq 0.60$, as for the circular unitary ensemble, whereas in the MBL phase we expect $r = (2\ln 2 - 1) \simeq 0.39$, the result for $\theta_n$ uncorrelated random numbers. In Fig. S1(a) we show $r$ for various $L$ with periodic boundary conditions, computed from averages over the entire spectrum and over $N = 10^4$ disorder realisations. There is a clear transition in behaviour, which becomes sharper with increasing $L$, and this result indicates a critical coupling $J_c \approx 0.07$. We have not attempted to locate the transition more precisely for this work, since we are concerned with behaviour deep in each phase rather than critical phenomena. The corresponding transition in the behaviour of the average SFF

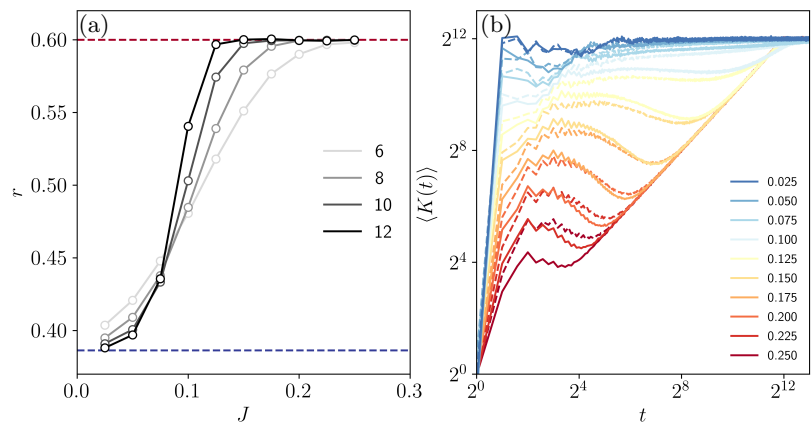

FIG. S1. Spectral statistics. (a) $r$-statistic for level correlations on the finest scales. The different shades show different system sizes $L$ (legend). The red dashed line is the result for Haar-random unitary matrices ($\approx 0.60$) and the blue is the result for uncorrelated levels ($2\ln 2 - 1 \approx 0.39$). Here we use periodic boundary conditions. (b) Average SFF $\langle K(t)\rangle$ for $L = 12$ and various $J$ (legend). The solid line shows the result with periodic boundary conditions and the dashed with open.

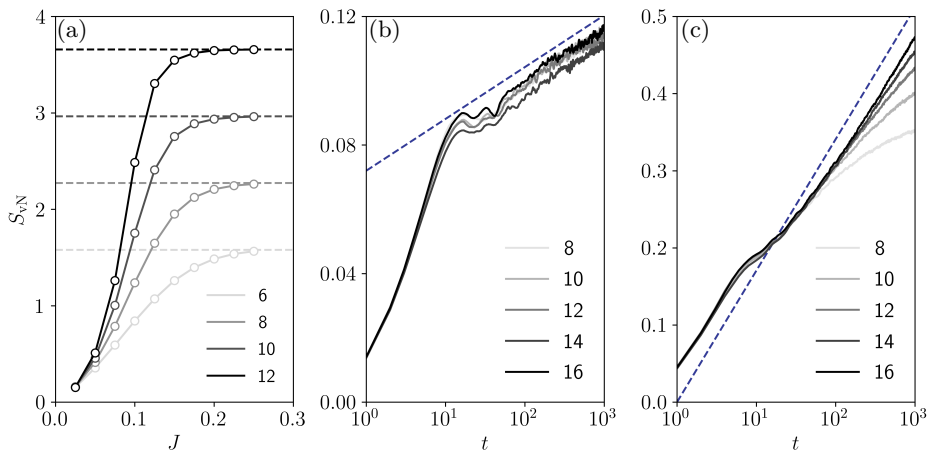

FIG. S2. Half-chain von Neumann entanglement entropy $S_{\mathrm{vN}}$. (a) $S_{\mathrm{vN}}$ for eigenstates of the Floquet operator as a function of $J$, and for various $L$ shown on the legend. The dashed lines show the Page value $(1/2)(L\ln 2 - 1)$ neglecting corrections of order $1/2^L$, and the statistical errors are not visible on this scale. (b) Growth of $S_{\mathrm{vN}}$ with time for circuits with $J = 0.025$, and starting from a randomly chosen eigenstate of the decoupled system. The legend shows $L$, and dashed lines indicate logarithmic growth. (c) As for (b), here with $J = 0.05$.

$\langle K(t)\rangle$ is shown in Fig. S1(b) for $L = 12$ sites, $1 \leq t \leq 2^{13}$, and both periodic and open boundary conditions. Here also $N = 10^4$.

The difference between the ergodic and MBL phases is also clear in eigenstate entanglement [S2], and in the growth of entanglement with time from initially unentangled states [S3–S6]. We focus on the half-chain von-Neumann entanglement entropy $S_{\mathrm{vN}} = -\mathrm{Tr}[\rho\ln\rho]$, where $\rho$ is the reduced density matrix over half of a chain with periodic boundary conditions. In Fig. S2(a) we show $S_{\mathrm{vN}}$ for Floquet eigenstates, averaged over the spectrum and $N = 10^4, 10^4, 10^3, 10^2$ disorder realisations for $L = 6, 8, 10, 12$, respectively. At large $J$ in the ergodic phase our results approach the Page value $(1/2)(L\ln 2 - 1) + O(2^{-L/2})$ corresponding to volume-law entanglement [S7]. For small $J$ on the other hand we find an $L$-independent $S_{\mathrm{vN}}$, which in one dimension corresponds to area-law entanglement [S2]. In Figs. S2(b) and (c) we calculate $S_{\mathrm{vN}}$ as a function of time $t$ starting from eigenstates of the decoupled system ($J = 0$). In the MBL phase it is expected that $S_{\mathrm{vN}}$ increases logarithmically with $t$ [S3–S6]. In Fig. S2(b) we evolve the states under Floquet operators with $J = 0.025$ and in (c) with $J = 0.05$, averaging over $10^4$ disorder realisations in each case, and find approximately this behaviour.

## S2. CONSTRUCTION OF THE TRANSFER MATRICES

In this section we give details on the transfer matrices which generate the SFF $K(t) = |\mathrm{Tr}W(t)|^2$, following Ref. [S8]. The evolution operator $W(t)$ is constructed from $4 \times 4$ unitary matrices $U_{x,x+1}$ acting on pairs of sites $(x, x+1)$. With $x$ even $U_{x,x+1}$ evolves the sites from integer time $r$ to $(r + 1/2)$. The amplitude for evolving the sites from the product state $|a_r\rangle \otimes |a'_r\rangle$ to $|b_r\rangle \otimes |b'_r\rangle$, where unprimed and primed labels correspond to $x$ and $(x+1)$, respectively, is $U_{b_r b'_r, a_r a'_r}$, where we have suppressed the position label $(x, x+1)$.

From these matrix elements we define a non-unitary matrix $\tilde{U}$ via $\tilde{U}_{a_r b_r, a'_r b'_r} = U_{b_r b'_r, a_r a'_r}$. $\tilde{U}$ acts on the path of site $(x+1)$ over the time interval $(r, r + 1/2)$. Taking the tensor product of $t$ copies of $\tilde{U}$ we find $\tilde{U}^{\otimes t}$, a non-unitary matrix which acts on the entire path, or orbit, of the site $(x+1)$. $\tilde{U}^{\otimes t}$ can be visualised as acting on a $2t$-site spin chain with periodic boundary conditions [see Fig. 1(c) of the main text].

Using the half-step time translation operator for orbits, $S$ [see main text], $\mathrm{Tr}W(t)$ is given by

$$\mathrm{Tr}W(t) = \mathrm{tr}[\tilde{U}^{\otimes t}_{0,1} S \tilde{U}^{\otimes t}_{1,2} S^{-1} \ldots \tilde{U}^{\otimes t}_{L-2,L-1} S \tilde{U}^{\otimes t}_{L-1,0} S^{-1}], \tag{S1}$$

with periodic boundary conditions, which necessitates $L$ even. The trace $\mathrm{tr}$ is over single-site orbits. Across bonds $(x, x+1)$ with $x$ even we act with $\tilde{U}^{\otimes t}_{x,x+1}$ and with $x$ odd we act instead with $S\tilde{U}^{\otimes t}_{x,x+1}S^{-1}$.

The matrices which generate the SFF in this way are tensor products of those generating $\mathrm{Tr}W(t)$ and its conjugate.

Defining $\mathcal{R}_{x,x+1} = \tilde{U}_{x,x+1}^{\otimes t} \otimes [\tilde{U}_{x,x+1}^{\otimes t}]^*$ we find, again with periodic boundary conditions,

$$K(t) = \text{tr}[\mathcal{R}_{0,1}\mathcal{S}\mathcal{R}_{1,2}\mathcal{S}^{-1}\ldots\mathcal{R}_{L-2,L-1}\mathcal{S}\mathcal{R}_{L-1,0}\mathcal{S}^{-1}], \tag{S2}$$

where $\mathcal{S} = S \otimes S$ is the half-step time-translation operator for orbit pairs. Time periodicity implies $[\mathcal{S}^2, \mathcal{R}_{x,x+1}] = 0$, and this allows us to write $K(t)$ in terms of transfer matrices which have the same structure on each bond, $\mathcal{S}\mathcal{R}_{x,x+1}$. With open boundary conditions for example,

$$K(t) = \langle \mathcal{B}_L|(\mathcal{S}\mathcal{R}_{0,1})\ldots(\mathcal{S}\mathcal{R}_{L-2,L-1})|\mathcal{B}_R\rangle, \tag{S3}$$

where $\langle\mathcal{B}_L|$ and $|\mathcal{B}_R\rangle$ are vectors invariant under $S^2 \otimes 1$ and $1 \otimes S^2$ which encode the boundary conditions [S8]. For example $|\mathcal{B}_R\rangle$ in Eq. (S3) is

$$|\mathcal{B}_R\rangle = \sum_{\substack{a_0 b_0 \ldots \\ a_0^* b_0^* \ldots}} \Big(\prod_{r=0}^{t-1} \delta_{a_{r+1}b_r}\delta_{a_{r+1}^*b_r^*}\Big)|a_0 b_0 \ldots\rangle \otimes |a_0^* b_0^* \ldots\rangle. \tag{S4}$$

The disorder average of each $\mathcal{R}_{x,x+1}$ is the same on each bond, $\langle\mathcal{R}_{x,x+1}\rangle = \langle\mathcal{R}\rangle$, so the average SFF $\langle K(t)\rangle$ can be expressed in terms of powers of the average transfer matrix $\mathcal{S}\langle\mathcal{R}\rangle$. In large systems the spectral statistics of the Floquet operator are therefore determined by the leading eigenvalues of $\mathcal{S}\langle\mathcal{R}\rangle$. As indicated in the main text, the corresponding left and right eigenvectors ($\langle\nu,t;L|$ and $|\nu,t;R\rangle$, respectively) are invariant under $\mathcal{S}^2$. From this property it is straightforward to derive a relation between them. The eigenvectors satisfy

$$\mathcal{S}\langle\mathcal{R}\rangle|\nu,t;R\rangle = \lambda(\nu,t)|\nu,t;R\rangle, \tag{S5}$$
$$\langle\nu,t;L|\mathcal{S}\langle\mathcal{R}\rangle = \lambda(\nu,t)\langle\nu,t;L|,$$

and the matrix $\langle\mathcal{R}\rangle$ is real and symmetric. Taking the Hermitian conjugate of the second equation, and then multiplying on the left by $\mathcal{S}$, we find

$$\mathcal{S}\langle\mathcal{R}\rangle\mathcal{S}|\nu,t;L\rangle = \lambda^*(\nu,t)\mathcal{S}|\nu,t;L\rangle, \tag{S6}$$

where we have used $\mathcal{S}^{-1}|\nu,t;L\rangle = \mathcal{S}|\nu,t;L\rangle$. Within each of the sectors labelled by $\nu$ the leading eigenvalue $\lambda(\nu,t)$ is real and non-degenerate, and consequently we must have $\mathcal{S}|\nu,t;L\rangle = |\nu,t;R\rangle$.

### S3. CORRELATIONS OF THE LOCAL ORDER PARAMETER

To express the SFF $K(t) = |\text{Tr}W(t)|^2$ as a sum over orbit pairs, we first choose a basis for the many-body space. As in the main text it is convenient to use product states over sites, and to be concrete we work in the basis of eigenstates of the Pauli-$Z$ operators $\{Z(x)\}$ for $x = 0\ldots(L-1)$. We can then write $\text{Tr}W(t) = \sum_P A_P$, where $P$ are closed paths of $t$ steps in the space of $Z^{\otimes L}$ eigenstates. The complex amplitudes $A_P$ are then products of $t$ matrix elements of $W$ in this basis, and the SFF is

$$K(t) = \sum_{PQ} A_P A_Q^*. \tag{S7}$$

The weights associated with each orbit pair $A_P A_Q^*$ are complex, although $K(t)$ is real and non-negative. In our work we are interested in averaged properties of the sum over orbit pairs in Eq. (S7).

In the main text we have introduced a clock-model order parameter $\mathcal{C}$ for single-site orbit pairs $|a_0 b_0 \ldots\rangle \otimes |a_0^* b_0^* \ldots\rangle$. There $\mathcal{C}$ has been chosen to be a diagonal operator acting as

$$\mathcal{C}|a_0 b_0 \ldots\rangle \otimes |a_0^* b_0^* \ldots\rangle = \frac{2}{t}\sum_{p,r}\delta_{a_p a_r^*}e^{2\pi i(p-r)/t}|a_0 b_0 \ldots\rangle \otimes |a_0^* b_0^* \ldots\rangle. \tag{S8}$$

In this way we can assign the many-body orbit pairs in Eq. (S7) values of the order parameter at each site $x$. For orbits $P, Q$ we write these numbers as $\mathcal{C}_{P,Q}(x)$. The one-point function is naturally defined as

$$\mathcal{O}_1(x,t) = \frac{\langle\sum_{PQ} A_P A_Q^* \mathcal{C}_{P,Q}(x)\rangle}{\langle\sum_{PQ} A_P A_Q^*\rangle}, \tag{S9}$$

where we have averaged the numerator and denominator separately so that $\mathcal{O}_1(x,t)$ probes the same average of the sum over orbit pairs as $\langle K(t) \rangle$.

To write $\mathcal{O}_1(x,t)$ in terms of the evolution operator we use $2\delta_{a_p a_r^*} = (Z_{a_p a_p}(x) Z_{a_r^* a_r^*}(x) + 1)$ and find

$$\mathcal{O}_1(x,t) \times \langle K(t) \rangle = \frac{1}{t} \sum_{p,r} e^{2\pi i (p-r)/t} \langle \mathrm{Tr}[W(t-p) Z(x) W(p)] \mathrm{Tr}[W(t-r) Z(x) W(r)]^* \rangle. \tag{S10}$$

Since $W(p)W(t-p) = W(t)$ for $p$ integer, the sum on $p$ in Eq. (S10) ensures that the one-point function $\mathcal{O}_1(x,t) = 0$.

The two-point function $\mathcal{O}_2(x,y,t)$ is given by inserting $\mathcal{C}^*$ at the site $x$ and $\mathcal{C}$ at the site $y$,

$$\mathcal{O}_2(x,y,t) = \frac{\langle \sum_{PQ} A_P A_Q^* \mathcal{C}_{P,Q}^*(x) \mathcal{C}_{P,Q}(y) \rangle}{\langle \sum_{PQ} A_P A_Q^* \rangle}. \tag{S11}$$

In this way the numerator is related to a scalar product of (two-component vector) local order parameters. To find an expression in terms of $W(t)$ we must sum over $p,r$ at each of the sites $x$ and $y$. One finds an expression of the form

$$\mathcal{O}_2(x,y,t) \times \langle K(t) \rangle = \frac{1}{t^2} \sum_{p_x, r_x, p_y, r_y = 0}^{t-1} e^{-2\pi i (p_x - r_x - p_y + r_y)} M(p_x, p_y) M^*(r_x, r_y), \tag{S12}$$

where $M(p_x, p_y)$ is related to a trace over $\mathrm{Tr} W(t)$ but with an insertion of $Z(x)$ at time step $p_x$, and $Z(y)$ at time step $p_y$. It can be shown that

$$\mathcal{O}_2(x,y,t) \times \langle K(t) \rangle = \langle | \sum_{r=0}^{t-1} \mathrm{Tr}[W(t-r) Z(x) W(r) Z(y)] e^{2\pi i r/t} |^2 \rangle. \tag{S13}$$

Writing $W(t) = \sum_n e^{i\theta_n t} |n\rangle\langle n|$ and evaluating the sum over $r$ in Eq. (S13), we find

$$\mathcal{O}_2(x,y,t) \times \langle K(t) \rangle = \langle | \sum_{mn} G_{mn}(t) \langle n | Z(x) | m \rangle \langle m | Z(y) | n \rangle |^2 \rangle, \tag{S14}$$

where the selection function $G_{mn}(t) = e^{i\theta_n t} \sum_{r=0}^{t-1} e^{i(\theta_m - \theta_n + 2\pi/t)r}$ emphasises contributions from pairs of levels separated in quasienergy by $(\theta_n - \theta_m) \sim 2\pi/t$.

As a toy model for the ergodic phase we briefly discuss the two-point function in the case where $W$ is a $2^L \times 2^L$ Haar-random unitary matrix. Then for $t \ll 2^L$ the right-hand side of Eq. (S13) can be calculated using standard techniques [S9]. Expanding the modulus-squared sum in the average in Eq. (S13) we have $t^2$ terms, but only the $t$ diagonal terms contribute for large $2^L$. Each of these gives unity, and because $\langle K(t) \rangle = t$ we find $\mathcal{O}_2(x,y,t) = 1$.

In addition to expressing $\mathcal{O}_2(x,y,t)$ in terms of the spectral decomposition of $W(t)$, it is useful to relate it to the transfer matrix that generates the SFF. This is straightforward because the numerator and denominator of Eq. (S11) are averaged independently. For example with periodic boundary conditions, and $x$ and $y$ both odd,

$$\mathcal{O}_2(x,y,t) = \frac{\mathrm{tr}[\mathcal{S}^{-L}(\mathcal{S}\langle\mathcal{R}\rangle)^{L-|x-y|}\mathcal{C}^*(\mathcal{S}\langle\mathcal{R}\rangle)^{|x-y|}\mathcal{C}]}{\mathrm{tr}[\mathcal{S}^{-L}(\mathcal{S}\langle\mathcal{R}\rangle)^L]}, \tag{S15}$$

where $|x-y|$ is defined modulo $L$. At any finite $t$ the eigenvalue $\lambda(0,t) > \lambda(\nu \neq 0, t)$, so taking the thermodynamic limit $L \to \infty$ with $t$ and $|x-y|$ fixed we find

$$\mathcal{O}_2(x,y,t) = [\lambda(1,t)/\lambda(0,t)]^{|x-y|} \langle 0,t;L|\mathcal{C}^*|1,t;R\rangle\langle 1,t;L|\mathcal{C}|0,t;R\rangle + \ldots, \tag{S16}$$

where the ellipses denote the contributions of subleading eigenvalues, which are suppressed for large $|x-y|$. Here we have used the selection rule coming from $(S^{-2} \otimes 1)\mathcal{C}(S^2 \otimes 1) = e^{-2\pi i/t}\mathcal{C}$. In the main text we have denoted the above amplitude $F(t) = \langle 0,t;L|\mathcal{C}^*|1,t;R\rangle\langle 1,t;L|\mathcal{C}|0,t;R\rangle$. For $x$ and $y$ both even we instead find $F(t) = \langle 0,t;L|\mathcal{SC}^*\mathcal{S}|1,t;R\rangle\langle 1,t;L|\mathcal{SCS}|0,t;R\rangle = \langle 1,t;L|\mathcal{C}|0,t;R\rangle^* \langle 0,t;L|\mathcal{C}^*|1,t;R\rangle^*$, where we have used $\mathcal{S}|\nu,t;L\rangle = |\nu,t;R\rangle$ [see Sec. S2]. Since the choice of spatial origin is arbitrary, this must be equal to the amplitude appearing in the case where $x$ and $y$ are both odd, and this implies that $F(t) = \langle 0,t;L|\mathcal{C}^*|1,t;R\rangle\langle 1,t;L|\mathcal{C}|0,t;R\rangle$ is real. When one of $x$ and $y$ is odd, and the other is even, the amplitude is instead $F(t) = |\langle 1,t;L|\mathcal{C}|0,t;R\rangle|^2$.

It is clear from the above that the two-point function decays over a lengthscale $\xi(t)$ set by the ratio $\lambda(1,t)/\lambda(0,t)$. To extract this it is convenient to consider a system with open boundary conditions, and with $x=0$ and $y=(L-1)$ the two end sites. We find

$$\mathcal{O}_2(0, L-1, t) = [\lambda(1,t)/\lambda(0,t)]^{L-1} \frac{\langle \mathcal{B}_L | \mathcal{C}^* | 1, t; R \rangle \langle 1, t; L | \mathcal{C} | \mathcal{B}_R \rangle}{\langle \mathcal{B}_L | 0, t; R \rangle \langle 0, t; L | \mathcal{B}_R \rangle} + \ldots \tag{S17}$$

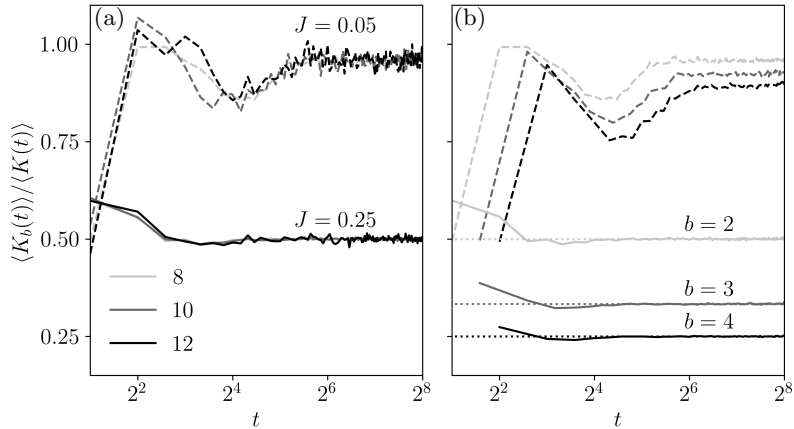

FIG. S3. Response of the average SFF to local symmetry-breaking fields in the MBL ($J = 0.05$, dashed) and ergodic ($J = 0.25$, solid) phases. (a) $\langle K_2(t)\rangle/\langle K(t)\rangle$ for various $L$ (legend). (b) $\langle K_b(t)\rangle/\langle K(t)\rangle$ for $L = 8$ and $b = 2, 3, 4$ as indicated. Horizontal dotted lines show $1/b$, the value expected in the symmetry-broken phase. $\langle K_b(t)\rangle$ is defined only for $t$ an integer multiple of $b$, and here we use periodic boundary conditions. For $L = 8, 10$ we average over $N = 10^5$ disorder realisations, and for $L = 12$ over $N = 10^4$.

for large $L$. By following the scaling of $\mathcal{O}_2(0, L-1, t)$ with $L$ at fixed $t$ we extract $\xi(t) = 1/\ln[\lambda(0,t)/\lambda(1,t)]$ as shown in Fig. 3(d) of the main text.

## S4. RESPONSE TO SYMMETRY-BREAKING FIELDS

In the main text, and in Sec. S3, we have studied correlations of the clock order parameter. In the ergodic phase we have shown that there are long-range correlations of this order parameter, as in the symmetry-broken phase of a classical clock model, whereas in the MBL phase the correlations are short ranged. Another way to investigate the sum over many-body orbit pairs generating the SFF is to consider the response to a symmetry-breaking field. For example at times $t$ with $t/b$ and $b$ both integers a natural choice is to break the symmetry locally from $t$-fold to $t/b$-fold. In the symmetry-broken phase of a classical clock model, one expects this to reduce the partition function by a factor $b$. In the symmetric phase there is a much weaker response.

Locally breaking the symmetry from $t$-fold to $t/b$-fold can be achieved in practice by locally punctuating the time evolution with $2 \times 2$ Haar-random unitary matrices which repeat only after $b$ time steps. For example with $b = 2$, breaking the symmetry at just one site $x$, the evolution operator for $t$ even is changed from $W(t) = W^t$ to $(W^{(2)}W^{(1)})^{t/2}$, where $W^{(1)}$ differs from $W^{(2)}$ only in the realisation of the randomness at $x$. Since $W^{(2)}W^{(1)}$ is simply another Floquet operator, but now for evolution with period 2, it is natural to expect that in the ergodic phase the new SFF for time $t$ (and so $t/2$ periods) is $t/2$. More generally we expect

$$\langle K_b(t)\rangle = \left\langle \left|\mathrm{Tr}[(W^{(b)}\ldots W^{(2)}W^{(1)})^{t/b}]\right|^2 \right\rangle \simeq t/b \tag{S18}$$

in the ergodic phase for large $t < 2^L$. In the limit of decoupled sites ($J = 0$) one instead finds that for $t \geq 2b$, $\langle K_b(t)\rangle = \langle K(t)\rangle$. For small $J < J_c$ in the MBL phase we therefore expect $\langle K_b(t)\rangle \simeq \langle K(t)\rangle$. In Fig. S3 we show $\langle K_b(t)\rangle/\langle K(t)\rangle$ as a function of $t$ for various $b$ and $L$, and find excellent agreement with the above predictions. There is a clear difference in response between the ergodic and MBL phases, demonstrating a form of long-range rigidity in the former that arises from the broken symmetry.

---



\end{document}